\documentclass[aps,prc,preprint,amsmath,showpacs,superscriptaddress,nofootinbib]{revtex4}
\usepackage{graphicx,epsfig,epstopdf}

\begin{document}

\title{Generalized equation of state for cold superfluid neutron stars}

\author{N.~Chamel}
\affiliation{Institut d'Astronomie et d'Astrophysique, CP-226, Universit\'e
Libre de Bruxelles, 1050 Brussels, Belgium}
\author{J. M. Pearson}
\affiliation{D\'epartement de Physique, Universit\'e de Montr\'eal, Montr\'eal (Qu\'ebec), H3C 3J7 Canada}
\author{S.~Goriely}
\affiliation{Institut d'Astronomie et d'Astrophysique, CP-226,
Universit\'e Libre de Bruxelles, 1050 Brussels, Belgium}

\date{\today}

\begin{abstract}
Mature neutron stars are expected to contain various kinds of superfluids in 
their interiors. Modeling such stars requires the knowledge of the mutual entrainment 
couplings between the different condensates. We present a unified equation of state 
describing the different regions of a neutron star with superfluid neutrons and superconducting 
protons in its core. 
\end{abstract}

\pacs{97.60.Jd,26.60.-c,26.60.Kp}

\maketitle


The existence of superfluids inside neutron stars is rather well-established. It was 
first suggested a long time ago~\cite{bog58} before the actual discovery of pulsars and
 was later supported by the observation of the long relaxation time following the first 
observed glitch in the Vela pulsar~\cite{baym69}. Since
then, glitches have been observed in many other pulsars. Superfluidity is also expected 
to play a key role in the origin of large glitches and several scenarios have been proposed, e.g. a 
catastrophic unpinning of superfluid vortices~\cite{and75}, a transition between turbulent and laminar 
superfluidity~\cite{per06} or a superfluid two-stream instability~\cite{kos09}.
Microscopic studies~\cite{dean03} indicate that the interior of mature neutron stars contain three kinds 
of condensates: a neutron superfluid permeating the inner layers of the crust and a mixture of a neutron 
superfluid with a proton superconductor in the core. The most massive neutron stars may contain additional 
superfluid species in their central cores such as hyperons and quarks. 
The critical temperatures below which these various transitions occur still remain very uncertain.

At finite temperatures, a superfluid is described by two velocity fields: one associated 
with the superflow and a normal fluid which carries all the entropy. This two-fluid model was first introduced 
by Tisza~\cite{tisza38} for explaining the unusual properties 
of superfluid helium $^4$He. With the discovery of superfluid $^3$He, it was  
realized that in superfluid mixtures such as $^3$He-$^4$He, the different condensates are coupled by (non-dissipative) 
mutual entrainment effects~\cite{ab75}: even though they can flow with their own velocity field, the momentum of 
each superfluid is not aligned with its corresponding velocity. Entrainment effects may have an important impact 
on the evolution of superfluid neutron stars (see e.g. Ref.~\cite{alp84}). 
B. Carter has developed an action principle for obtaining 
the relativistic hydrodynamic equations of superfluids and superconductors, as found in the interior 
of neutron stars~\cite{car01}. The superfluid mixture is described by a Lagrangian density $\Lambda$ which depends on 
the 4-current vectors $n_{_{\rm X}}^{\, \mu}$ of the different constituents labeled by X ($\mu$,$\nu$ are used to denote spacetime indices). 
Considering variations of the fluid particle trajectories, the action principle leads to the hydrodynamic equations 
\begin{equation}
\label{1}
n_{_{\rm X}}^{\, \mu}\varpi^{_{\rm X}}_{\!\mu\nu} + \pi^{_{\rm X}}_{\, \nu}\nabla_\mu n_{_{\rm X}}^{\, \mu}  = f^{_{\rm X}}_{\,\nu} \, ,
\end{equation}
where $f^{_{\rm X}}_{\,\nu}$ is 4-force density covector acting on the fluid $X$ and 
the vorticity 2-form $\varpi^{_{\rm X}}_{\!\mu\nu}\equiv \nabla_{\!\mu}\pi^{_{\rm X}}_{\,\nu} - \nabla_{\!\nu}\pi^{_{\rm X}}_{\,\mu}$ 
is defined as the exterior derivative of the 4-momentum covector
\begin{equation}
\label{2}
\pi^{_{\rm X}}_{\, \mu} = \frac{\partial \Lambda}{\partial n_{_{\rm X}}^{\, \mu}} =g_{\mu\nu}(\mathcal{B}^{_{\rm X}} n_{_{\rm X}}^{\, \nu} + \sum_{_{\rm Y\neq X}}\mathcal{A}^{_{\rm XY}}n_{_{\rm Y}}^{\, \nu})\, .
\end{equation}
The ``anomalous'' coefficients $\mathcal{A}^{_{\rm XY}}$ above arise from entrainment effects. 
This variational formalism has been adapted to non-relativistic superfluids using a 4D fully 
covariant framework~\cite{cc} in order to facilitate the matching between macroscopic and microscopic models, and also because 
Newtonian models have been widely used for qualitative studies of superfluid neutron stars. 
The relation between relativistic and non-relativistic models has been discussed in Ref.~\cite{cha08}. 

The electrically charged particles inside neutron stars are locked together by the interior magnetic field and co-rotate 
on very long time scales of the order of the age of the star~\cite{eas79}. 
In contrast the neutron superfluid being electrically 
uncharged can rotate at a different rate. This naturally leads to considering the interior of old neutron stars as a two-fluid 
mixture. However, as mentioned earlier the description of young and massive neutron stars may require more elaborate models~\cite{gus09}. 
The simplest model of cold superfluid neutron stars thus consists of a mixture of a neutron superfluid with a plasma of charged 
particles (superconducting protons and leptons in the core, nuclei and electrons in the crust). 
The core is assumed to be entirely 
superfluid and elasticity effects of the solid crust are neglected. The neutron and ``proton'' fluids are described by their 4-current 
vectors $n_n^\mu$ and $n_p^\mu$ respectively. Introducing the particle densities 
$n_{_{\rm X}}^2 c^2=-g_{\mu\nu}n_{_{\rm X}}^\mu n_{_{\rm X}}^\nu$ and using the notation $x^2 c^2=-g_{\mu\nu}n_n^\mu n_p^\nu$, 
the Lagrangian density of the two fluids can be expressed to lowest order in the relative currents as~\cite{cha08} 
\begin{equation}
\label{4}
\Lambda(n_n^\mu,n_p^\nu)=\lambda_0(n_n,n_p)+\lambda_1(n_n,n_p)(x^2-n_n n_p)\, .
\end{equation}
The first term $\lambda_0=-\mathcal{E}$ is related to the internal energy density $\mathcal{E}$
in the absence of relative currents, while the second term accounts for entrainment effects. 

\begin{figure}
\label{fig1}
\includegraphics[height=.27\textheight]{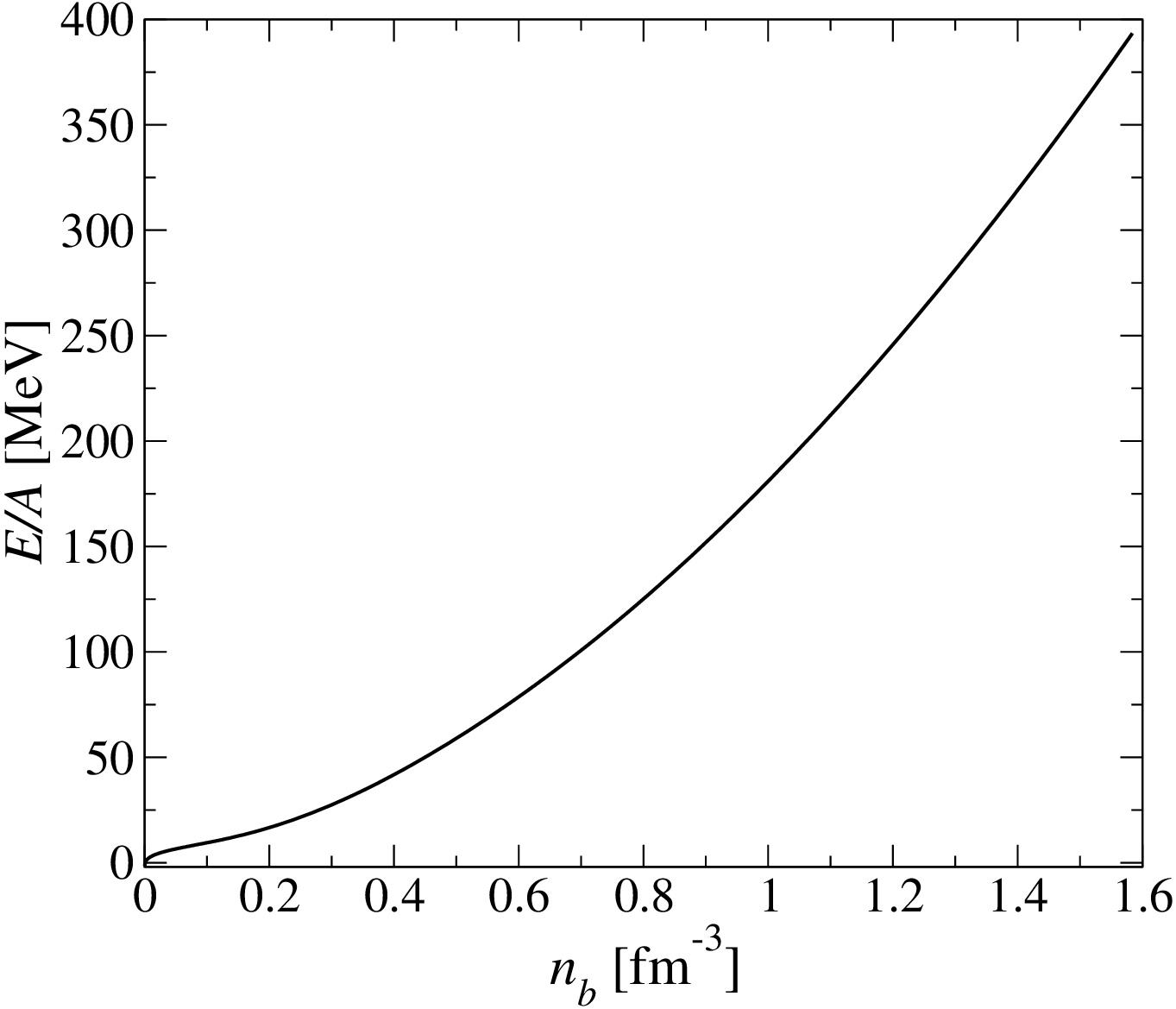}
\includegraphics[height=.27\textheight]{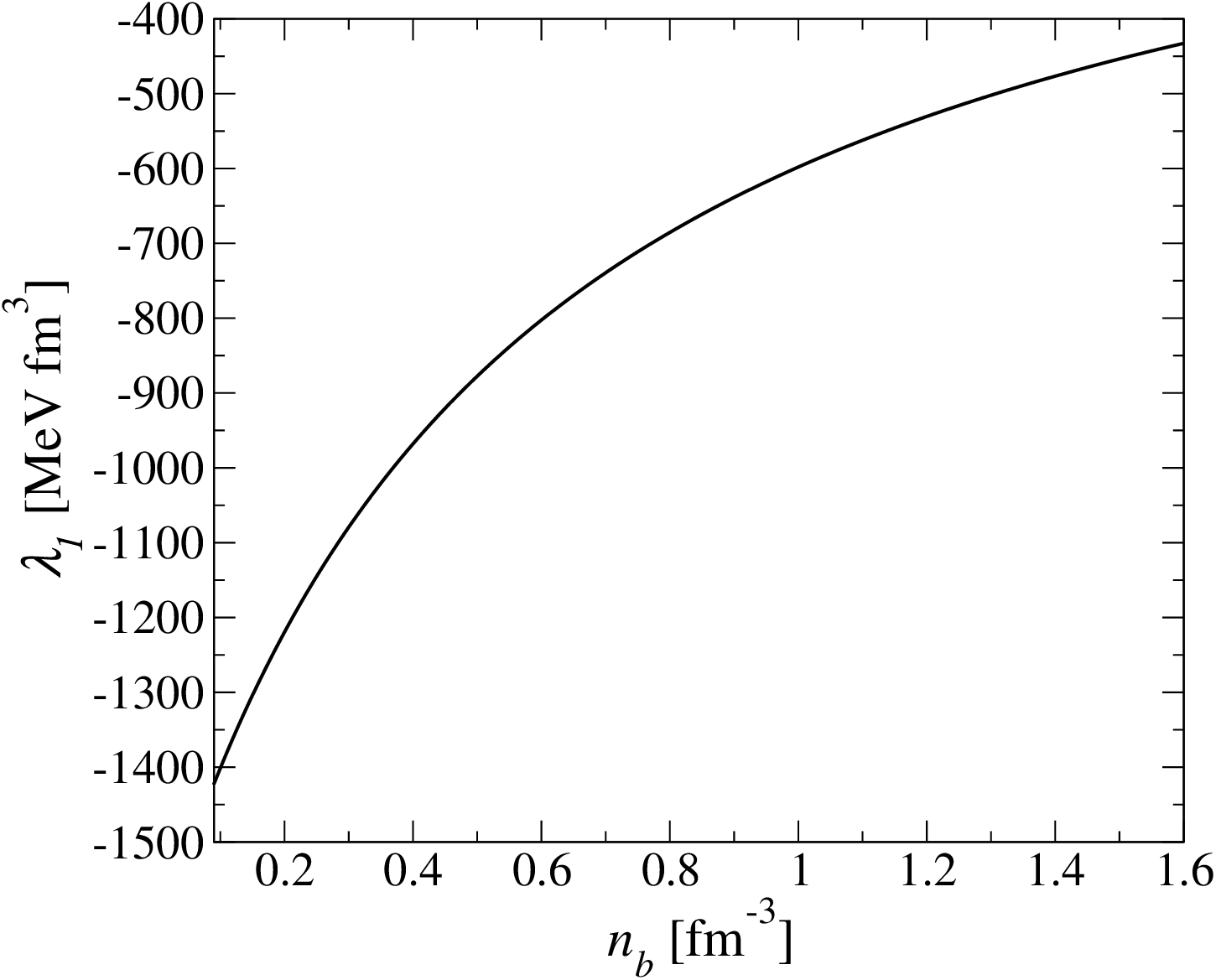}
\caption{Generalized equation of state of superfluid neutron stars, based on the 
HFB-17 mass model: binding energy per nucleon $E/A=-\lambda_0/n_b-M$ where $M$ is the nucleon mass 
(left panel) and entrainment coefficient $\lambda_1$ (right panel) as a function of the baryon density $n_b$.
}
\end{figure}

\begin{figure}
\label{fig2}
\includegraphics[height=.27\textheight]{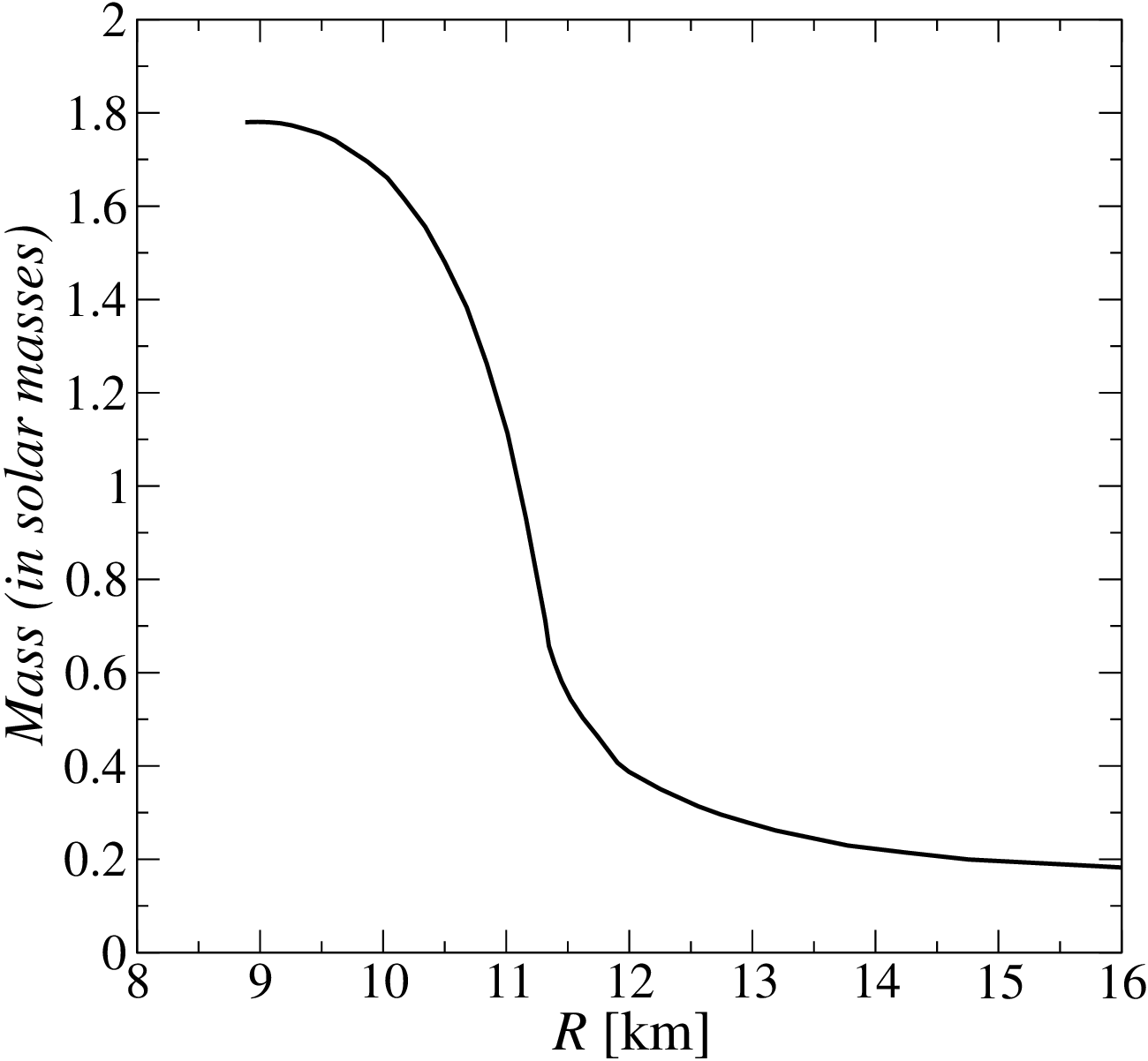}
\includegraphics[height=.27\textheight]{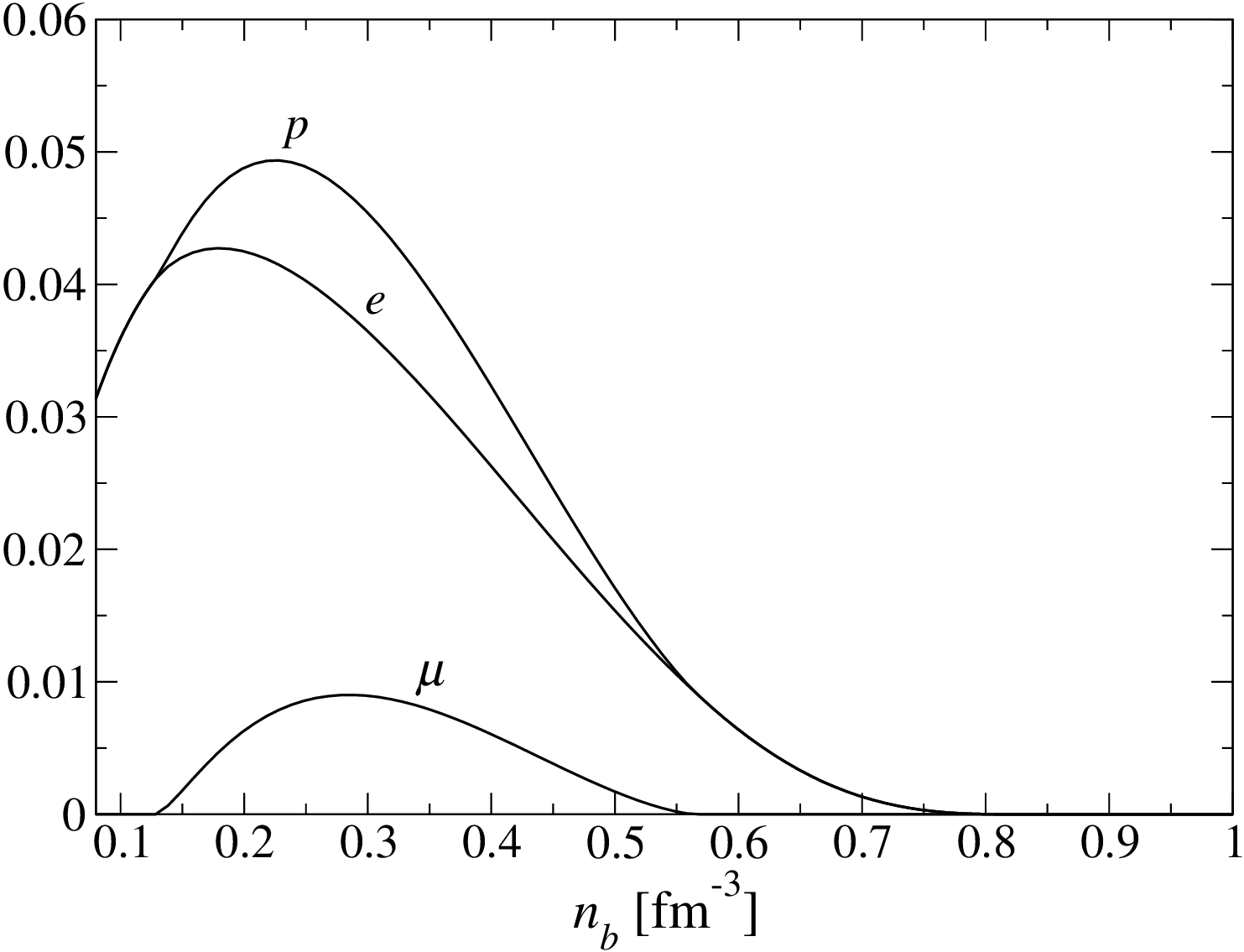}
\caption{Left panel: Mass-radius relation for non-rotating neutron stars using the unified equation of state 
shown in Fig.~\ref{fig1}. Right panel: fractions $n_{_{\rm X}}/n_b$ of protons (${\rm X}=p$), electrons (${\rm X}=e$) and muons (${\rm X}=\mu$).}
\end{figure}

We have calculated the functions $\lambda_0(n_n,n_p)$ and $\lambda_1(n_n,n_p)$ for all regions of a neutron star using
the self-consistent mean-field method with Skyrme effective interactions~\cite{sto07}. 
This method has been very successful in 
describing the structure and the dynamics of medium-mass and heavy nuclei and has been also widely 
applied to the description of neutron stars and supernova cores. 
This method allows for a unified treatment of both 
homogeneous and inhomogeneous matter with a reduced computational cost.
In this work, we have used the Skyrme interaction BSk17 underlying our Hartree-Fock-Bogoliubov (HFB) nuclear mass model HFB-17~\cite{gcp09}. 
This model fits essentially all the available experimental data with rms deviations of 0.58 MeV thus ensuring that nuclei in the crust
of a neutron star will be properly described. Moreover this model was constrained to reproduce various properties of infinite 
homogeneous nuclear matter as obtained from many-body calculations using realistic nucleon-nucleon potentials. In particular, 
this model fits a realistic equation of state of neutron matter and can therefore be also reliably applied to describe the 
liquid core of neutron stars. We have determined the equilibrium structure and equation of state of the outer crust of neutron stars 
following the standard approach of Ref.~\cite{bps71}. In this region, the only microscopic inputs are nuclear masses. We have used 
the HFB-17 nuclear mass table or experimental data when available. Results can be found in Ref.~\cite{pear09}. 
For the inner crust, we have applied the fourth-order Extended Thomas-Fermi method with proton shell effects added via the 
Strutinsky-Integral theorem~\cite{onsi08}. We have found that above the density $\rho\simeq 1.4\times 10^{14}$ g~cm$^{-3}$,
the crust dissoves into a uniform plasma of neutrons, protons and electrons. Muons appear at densities above $\rho\simeq 2.1\times 10^{14}$ 
g~cm$^{-3}$. Leptons are treated as relativistic Fermi gases. The functions $\lambda_0(n_n,n_p)$ and $\lambda_1(n_n,n_p)$ in the 
ground-state of cold dense matter are shown in Fig.~\ref{fig1} (analytic expressions for the core can be found in Ref.~\cite{cha08}). 
The mass-radius relation of non-rotating neutron stars (obtained after solving the Tolman-Oppenheimer-Volkoff equations) and the 
composition of their core are shown in Fig.~\ref{fig2}. Causality is satisfied inside any stable neutron star. 

Even though entrainment effects in neutron star cores have been studied for a long time~\cite{sauls89}, it has been 
only recently realized that similar effects should also occur in the inner crust~\cite{carter06}. Taking into account the 
neutron superfluid in the crust is of prime importance for modeling pulsar glitches~\cite{cha06} or quasi-periodic oscillations in 
Soft-Gamma Repeaters~\cite{pet10}. Calculations of $\lambda_1$ in the crust region will be reported elsewhere. 

{\it Acknowledgments}. 

This work was financially supported by the FNRS (Belgium), the Communaut\'e 
fran\c{c}aise de Belgique (Actions de Recherche Concert\'ees), the NSERC (Canada) and 
by CompStar, a Research Networking Programme of the European Science Foundation.

\end{document}